\documentclass[english,11pt,draftcls,onecolumn]{IEEEtran}
\usepackage[T1]{fontenc}
\usepackage[latin9]{inputenc}
\usepackage{color}
\usepackage{babel}
\usepackage{amsmath}
\usepackage{amssymb}
\usepackage{graphicx}
\usepackage[unicode=true,
 bookmarks=true,bookmarksnumbered=true,bookmarksopen=true,bookmarksopenlevel=1,
 breaklinks=false,pdfborder={0 0 0},backref=false,colorlinks=false]
 {hyperref}
\hypersetup{pdftitle={Your Title},
 pdfauthor={Your Name},
 pdfpagelayout=OneColumn,pdfnewwindow=true,pdfstartview=XYZ,plainpages=false}

\makeatletter
\usepackage{babel}

\usepackage{babel}

\usepackage{babel}
\ifCLASSOPTIONcompsoc
\else
\fi

\makeatother

\begin{document}

\title{Sum-rate maximization of OFDMA femtocell networks that incorporates
the QoS of macro mobile stations}

\author{\authorblockN{ Chunguo Li} }

\author{Chunguo Li, Stanford University, CA \vspace{-25pt}}
\maketitle
\begin{abstract}
This paper proposes a power allocation scheme with co-channel allocation
for a femto base station (BS) that maximizes the sum-rate of its own
femto mobile stations (MSs) with a constraint that limits the degradation
of quality of service (QoS) of macro MSs. We have found a closed-form
solution for the upper limit on the transmission power of each sub-channel
that satisfies the constraint in a probabilistic sense. The proposed
scheme is practical since it uses only the information easily obtained
by the femto BS. Moreover, our scheme meets the constraint with minimal
degradation compared to the optimal sum-rate of the femto MSs achieved
without the constraint. \end{abstract}

\section{Introduction }

Femtocell technology has recently attracted significant attention
as a way to enhance the performance of wireless cellular systems for
indoor areas \cite{Femtoforum,Chandrasekhar-COMMAG08}. Because of
their inherent low-power, short-range characteristics, a number of
femto base stations (BSs) can be deployed within a macrocell to achieve
a high degree of spatial reuse of the spectrum resources. Moreover,
there is no cost associated with site acquisition and backhaul connectivity
for femto BSs (F-BSs) because F-BSs are installed in customer premises
that have broadband access.

When femtocells and macrocells share the same frequency resources,
mutual interference, referred to as cross-tier interference, is present.
In this study, we considered the situation in which a macro BS (M-BS)
always transmits with the same power across sub-channels and does
not change its operation after the introduction of the femtocells.
Because the transmission power used by a M-BS is high compared to
the low transmission power used by a F-BS, when the femtocell is located
near the M-BS, the signal strength received by a femto mobile station
(F-MS) from the M-BS is much stronger than that from the F-BS. In
this case, different frequency resources should be allocated to the
femtocell to avoid the severe interference caused by the M-BS \cite{Guvenc};
this allocation is referred to as orthogonal channel allocation \cite{Chandrasekhar-TCOM09}.

In practice, femtocells are more likely to be installed in outer regions
of a macrocell to overcome the weak signal strength received by MSs
from the M-BS in those areas. In those regions, the signal strength
received by a M-MS from a F-BS is relatively strong given the poor
performance of the M-MS. In this case, if the F-BS appropriately controls
its transmit power, the femtocell and the macrocell can efficiently
share the same frequency resources; this allocation is referred to
as co-channel allocation \cite{Chandrasekhar-TWC09}. In terms of
area spectral efficiency \cite{Alouini}, co-channel allocation has
potential gains over orthogonal allocation due to the additional level
of spatial reusability of frequency resources \cite{Sundaresan}.
With co-channel allocation, we propose a power allocation algorithm
for a F-BS that probabilistically limits the quality of service (QoS)
degradation of M-MSs caused by the cross-tier interference from the
F-BS. Notice here, central coordinating across the F-BS and the M-BS
is another direction to pursue, which lies in the framework of coordinate
multipoint (CoMP) \cite{fan:comp0,fan:comp1,fan:comp2}. However, we will not use the CoMP technique in this work.

In this study, we considered an orthogonal frequency division multiple
access (OFDMA) system that is a basis for IEEE802.16e WiMAX \cite{WiMAX}
and 3GPP long-term evolution (LTE) \cite{LTE2}. In OFDMA systems,
total bandwidth is divided into several sub-channels; hence, a F-BS
can adjust the transmit power of each sub-channel to maximize its
performance. The cross-tier interference problem for downlink OFDMA
femto/macro two-tier networks has been investigated in our earlier
work \cite{Kim}. In that paper, we proposed a power allocation algorithm
for the F-BS that maximizes the weighted sum-rate of F-MSs and M-MSs
near the femtocell. In this one, we propose a power allocation algorithm
for a F-BS that maximizes the sum-rate of the F-MSs while considering
the QoS of each M-MS.

\textcolor{black}{After describing the system model under consideration
in Section II, we present the proposed femtocell power allocation
scheme in Section III. Section IV evaluates the performance gains
of the proposed scheme by simulation, and Section V concludes the
paper.}

\section{System Overview\label{Section 2} }

We considered the situation in which only one F-BS provides significant
interference to any M-MS; this is the case when F-BSs are sparsely
distributed in space within the macrocell. To limit the degradation
in QoS of the M-MS caused by the F-BS, the F-BS should know which
sub-channel is used to serve the M-MS located in the vicinity of the
femtocell, and the level of interference it could cause to the M-MS
if it were to use the same sub-channel. We consider the case in which
the F-BS periodically decodes the resource allocation information
(RAI) transmitted by the M-BS; the RAI is transmitted in the MAP field
of the downlink sub-frame in WiMAX \cite{WiMAX} and the physical
downlink control channel (PDCCH)) in LTE \cite{fan:lte2,LTE2}. Note
that when the F-BS receives and decodes the RAI in a downlink sub-frame,
the sub-frame must not be used by the F-BS to transmit data to its
F-MSs. If the F-BS receives and decodes the RAI frequently, for example
at every other frame, then the performance of the femtocell is degraded
by 50\%. On the other hand, if the interval between two readings of
the RAI is too long, then the RAI is likely to be outdated. With an
appropriate decoding interval, for instance once every ten frames,
the femtocell can maintain 90\% of its sum-rate; this degradation
is acceptable to femtocells whose sum-rates are generally very high.
In practice, stationary or slowly moving M-MSs are likely to be given
the same sub-channel over the multiple frames when it has traffic
to be downloaded from the M-BS since the channel conditions do not
change quickly for those M-MSs. For example, with the center frequency
of 2.5GHz, the coherence time of the channel for a MS which is moving
4km/h is 100msec, which corresponds to 100 sub-frames in LTE. Hence,
the RAI is effective with the appropriate interval. If the M-BS changes
its allocation of sub-channels, then the F-BS can track the allocation
within half of the interval on average. In this study, we assumed
that the resource allocation when the F-BS transmits is the same as
that decoded in the RAI.

In order to identify the level of interference that the F-BS could
cause M-MS $j$, the F-BS needs to know the channel gain of sub-channel
$n$ from the F-BS to M-MS $j$, which we denote by $H_{F,M_{j}}^{(n)}$.
However, in practice, the gain cannot be obtained by the F-BS, since
there is no direct feedback channel from the M-MS to the F-BS. Instead,
the F-BS can obtain the average of the channel gain over all sub-channels
denoted by $\overline{H}_{F,M_{j}}$ through the wired backhaul, since
M-MS $j$ reports the average received power from each adjacent BS
to its serving M-BS for specific purposes, such as hand-over; in LTE,
for example, the average received power is calculated from the primary
and secondary synchronization signals (PSS and SSS) that are transmitted
from each BS at every 5 msec \cite{Fan:lte}. A M-MS can differentiate
the source of the received signal since each BS uses a unique scrambling
code. Note that PSS and SSS are scrambled over the frequency range;
hence, M-MS $j$ still cannot measure $H_{F,M_{j}}^{(n)}$.

\begin{figure}[tbh]
\vspace{-10pt}

\centering \includegraphics[scale=0.45]{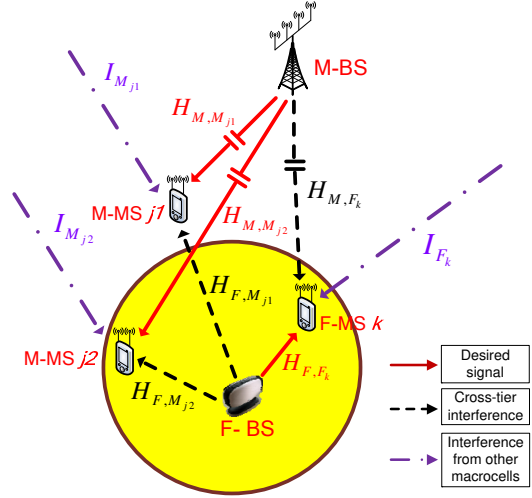}

\vspace{-10pt}

\centering\caption{Femto system model}

\vspace{-10pt}

\label{fig:environment}
\end{figure}

For simplicity, the building where the F-BS is installed is assumed
to be circular, as represented in Fig. \ref{fig:environment}. With
regard to the access policy of the femtocell, the closed policy is
considered; hence, a M-MS can be located inside the building as well
as outside. From the received signal strength of the uplink control
signal from M-MSs, the F-BS can also determine whether a specific
M-MS is inside or outside the building, given the additional attenuation
caused by the wall. The wall-loss is defined as the ratio of the strength
of the signal before penetrating the wall to that after penetrating
the wall, and is denoted as $L_{W}$. For fast-fading, we assume Rayleigh
fading.

\section{Femtocell Power Allocation}

\subsection{The Objective of a Femtocell}

\label{subsec:objective} In this study, the objective of a femtocell
is to maximize its sum-rate with an additional QoS constraint. The
QoS constraint is defined as
\begin{eqnarray}
Prob(\psi^{(n)}\leq\gamma^{(n)})\leq\epsilon^{(n)},~~\forall n,\label{ratio}
\end{eqnarray}
where $\psi^{(n)}$ is the ratio of the SINR with the femtocell to
that without the femtocell for the M-MS that uses sub-channel $n$,
$\gamma^{(n)}$ is the desired limit for the QoS degradation, and
$\epsilon^{(n)}$ is the allowable error that does not meet the limit.
It is impossible for the F-BS to guarantee the absolute QoS of the
M-MS, since the QoS of the M-MS could be below the required QoS even
before the introduction of the femtocell. Instead, the F-BS limits
the SINR degradation of the M-MS caused by the femtocell since a SINR
is a major value to determine QoS. Moreover, because of fast fading,
the QoS constraint should be expressed in a probabilistic sense. Note
that each M-MS may have a different QoS requirement depending on its
traffic type. In addition, for low SINR regions where F-BSs are normally
installed, the ratio in rate is approximated as that in the SINR,
since $log_{2}(1+SINR)$ is approximated as $SINR\times log_{2}e$.

With the QoS constraint, the objective of a femtocell is represented
as
\begin{eqnarray}
 &  & \underset{\left\{ p_{F}^{(1)},...,p_{F}^{(N)}\right\} }{\arg}\max\sum_{n=1}^{N}\left(C_{F}(p_{F}^{(n)})\right)\nonumber \\
 &  & ~~\textit{s.t.}~Prob(\psi^{(n)}\leq\gamma^{(n)})\leq\epsilon^{(n)},\nonumber \\
 &  & ~~~p_{F}^{(n)}\geq0,~\forall n,~\sum_{n=1}^{N}p_{F}^{(n)}\leq P_{F},
\end{eqnarray}
where $p_{F}^{(n)}$ is the transmit powers of the F-BS assigned to
sub-channel $n$, $P_{F}$ is the total transmit power, and $N$ is
the number of sub-channels. The rate of F-MS $k$ that uses sub-channel
$n$, $C_{F}(p_{F}^{(n)})$, is expressed as
\begin{eqnarray}
 &  & C_{F}(p_{F}^{(n)})\nonumber \\
= &  & log_{2}\left(1+\frac{p_{F}^{(n)}A_{F}H_{F,F_{k}}^{(n)}}{p_{M}^{(n)}A_{M}H_{M,F_{k}}^{(n)}/L_{W}+I_{F_{k}}^{(n)}+\sigma^{2}}\right),
\end{eqnarray}
where $p_{M}^{(n)}$ is the transmit powers of the M-BS assigned to
sub-channel $n$, $\sigma^{2}$ is the noise variance, and $A_{F}$
and $A_{M}$ are the antenna gains of a F-BS and a M-BS, respectively.
In the above equation, for sub-channel $n$, the channel gain from
the F-BS to F-MS $k$, the channel gain from the M-BS to F-MS $k$,
and the interference from other M-BSs to F-MS $k$ are represented
as $H_{F,F_{k}}^{(n)}$, $H_{M,F_{k}}^{(n)}$, and $I_{F_{k}}^{(n)}$,
respectively.

When M-MS $j$ located outside the building uses sub-channel $n$,
$\psi^{(n)}$ is further represented as
\begin{eqnarray}
\psi^{(n)} & \triangleq & \frac{SINR_{M_{j}}^{w/F,(n)}}{SINR_{M_{j}}^{w/o~F,(n)}}\nonumber \\
 & = & \frac{(p_{M}^{(n)}A_{M}H_{M,M_{j}}^{(n)})/\left(\frac{p_{F}^{(n)}A_{F}H_{F,M_{j}}^{(n)}}{L_{W}}+I_{M_{j}}^{(n)}+\sigma^{2}\right)}{(p_{M}^{(n)}A_{M}H_{M,M_{j}}^{(n)})/(I_{M_{j}}^{(n)}+\sigma^{2})}\nonumber \\
 & \simeq & \left(\frac{p_{F}^{(n)}A_{F}H_{F,M_{j}}^{(n)}}{I_{M_{j}}^{(n)}L_{W}}+1\right)^{-1},
\end{eqnarray}
where, for sub-channel $n$, $H_{M,M_{j}}^{(n)}$ is the channel gain
from the M-BS to M-MS $j$ and $I_{M_{j}}^{(n)}$ is the interference
from other M-BSs to M-MS $j$. Note that the approximation in the
last row is valid, since $I_{M_{j}}$ is normally much larger than
$\sigma^{2}$.

\subsection{The Power Limit on Each Sub-Channel}

\label{subsec:powerlimit} The power limit on each sub-channel is
derived from the QoS constraint. In the derivation, $I_{M_{j}}^{(n)}$
is divided into two parts: $\overline{I}_{M_{j}}$ for path loss and
shadowing and $i_{M_{j}}^{(n)}$ for fast fading. Similarly, $H_{F,M_{j}}^{(n)}$
is divided into $\overline{H}_{F,M_{j}}$ and $h_{F,M_{j}}^{(n)}$.
Note that $\overline{I}_{M_{j}}$ and $\overline{H}_{F,M_{j}}$ are
the same across different sub-channels, respectively. The power limits
are derived as
\begin{eqnarray}
 &  & Prob(\psi^{(n)}\leq\gamma^{(n)})\leq\epsilon^{(n)}\nonumber \\
 & \Leftrightarrow & Prob\left(p_{F}^{(n)}\geq\frac{L_{W}}{A_{F}}\frac{\overline{I}_{M_{j}}\times i_{M_{j}}^{(n)}}{\overline{H}_{F,M_{j}}\times h_{F,M_{j}}^{(n)}}\zeta^{(n)}\right)\leq\epsilon^{(n)}\nonumber \\
 & \Leftrightarrow & Prob\left(x\geq\frac{\kappa^{(n)}}{p_{F}^{(n)}}\right)\leq\epsilon^{(n)}\nonumber \\
 & \Leftrightarrow & 1-F_{X}\left(\frac{\kappa^{(n)}}{p_{F}^{(n)}}\right)\leq\epsilon^{(n)}\nonumber \\
 & \Leftrightarrow & \frac{1}{1+\kappa^{(n)}/p_{F}^{(n)}}\leq\epsilon^{(n)}~~~[\text{see APPENDIX for}~F_{X}]\nonumber \\
 & \Leftrightarrow & p_{F}^{(n)}\leq\kappa^{(n)}/\delta^{(n)}\triangleq K^{(n)},
\end{eqnarray}
where $\zeta^{(n)}\triangleq\frac{1}{\gamma^{(n)}}-1$, $x\triangleq\frac{h_{F,M_{j}}^{(n)}}{i_{M_{j}}^{(n)}}$,
$\kappa^{(n)}\triangleq\frac{L_{W}}{A_{F}}\frac{\overline{I}_{M_{j}}}{\overline{H}_{F,M_{j}}}~\zeta^{(n)}$,
and $\delta^{(n)}\triangleq\frac{1}{\epsilon^{(n)}}-1$.

F-BSs can know $\overline{H}_{F,M_{j}}/L_{W}$ from the uplink control
signal of M-MS $j$. Note that $L_{W}$ is not used for inside M-MSs.
$\zeta^{(n)}$ and $\delta^{(n)}$ are design parameters, and $A_{F}$
is a determined value. The only term that F-BS cannot know is $\overline{I}_{M_{j}}$.
In the proposed scheme, we approximate $\overline{I}_{M_{j}}$ to
$\overline{I}_{F}$, the interference that the F-BS experiences from
neighboring M-BSs, since the distance between the F-BS and M-MS $j$
is relatively very short compared to that between the neighboring
M-MSs and the F-BS or M-MS $j$. The approximation error could be
large when the distance between the F-BS and M-MS $j$ is long. Fortunately,
this is the case when the power limit $K^{(n)}$ is also very large.
In other words, $p_{Fopt}^{(n)}$ hardly approaches $K^{(n)}$ in
that case.

\subsection{Optimization}

\label{subsec:optimize} The optimization (2) with the QoS constraint
is a convex problem; therefore, the optimal solution can be found
from the Karush-Kuhn-Tucker (KKT) condition \cite{Boyd} as follows.
Let $S^{(n)}\triangleq p_{M}^{(n)}A_{M}H_{M,F_{k}}^{(n)}/L_{W}+I_{F_{k}}^{(n)}+\sigma^{2}$.
Then, the Lagrangian is $-log_{2}(1+\frac{p_{F}^{(n)}A_{F}H_{F,F_{k}}^{(n)}}{S^{(n)}})+\lambda(\sum_{n=1}^{N}p_{F}^{(n)}-P_{F})-\mu^{(n)}p_{F}^{(n)}+\nu^{(n)}(p_{F}^{(n)}-K^{(n)})$.
From the KKT conditions, the optimal solution is found to be
\begin{eqnarray}
p_{Fopt}^{(n)}=\min\left(K^{(n)},\max\left(0,\frac{1}{\lambda}-\frac{S^{(n)}}{\left(H_{F,F_{k}}^{(n)}A_{F}\right)}\right)\right)\nonumber \\
\text{[see APPENDIX for details].}
\end{eqnarray}

The above solution can be implemented as follows. First, the water-filling
(WF) algorithm \cite{Cover} is run, then the sub-channels that violate
$K^{(n)}$ are checked. Next, for those sub-channels, $p_{F}^{(n)}=K^{(n)}$
are assigned and the total power is reduced by $K^{(n)}$ per each
sub-channel. The above procedure is repeated for the remaining sub-channels
with the reduced total power until there is no sub-channel that violates
the upper limit.

\section{Numerical Results\label{Sec:sim} }

The Monte Carlo method was used to evaluate the performance of the
proposed scheme in a case in which 50 M-MSs are randomly located anywhere
in the macrocell including inside the building and 50 sub-channels
are allocated in a round-robin way to the M-MSs. To calculate interference
from neighboring M-BSs, we considered the adjacent 18 macrocells.

For the femtocell, without loss of generality, we randomly generated
a F-MS inside the building and assume that it can access all 50 sub-channels.
For path loss models, the IMT-Advanced Indoor Hotspot NLoS model and
the Urban Micro NLoS model with hexagonal cell layout \cite{IMT},
with the center frequency of 2.5GHz, were used for the femtocell and
the macrocells, respectively. The wall-loss is set as 3dB, which represents
the case in which a building is enclosed by windows or thin walls
\cite{wall-loss}. We set $R_{f}$, the radius of the building, to
30m and the radius of the macrocell to 500m. Antenna gains of a M-BS
and a F-BS are set as 15dBi and 2dBi, respectively.

\begin{figure}[tbh]
\centering \includegraphics[scale=0.3]{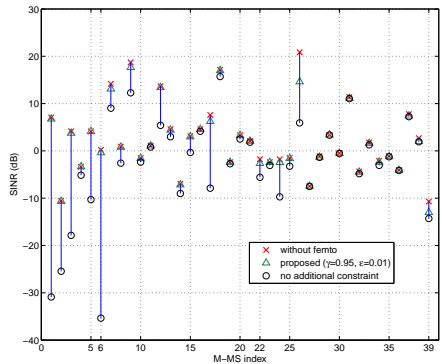}

\caption{SINRs of M-MSs located inside $3R_{f}$ ($d_{f}=400m$).}

\label{fig:MMS_abs}
\end{figure}

To more effectively represent the benefit of the proposed scheme,
we depict the SINR values of M-MSs located inside $3R_{f}$ in Fig.
\ref{fig:MMS_abs}. To obtain the figure, we repeated the simulation
20 times; hence, 1,000 M-MS were generated. Indices from 1 to 6 represent
the M-MSs located inside the building. Similarly, indices from 7 to
22 and from 23 to 39 correspond to the M-MS located between $R_{f}$
and $2R_{f}$ and between $2R_{f}$ and $3R_{f}$, respectively. Fig.
\ref{fig:MMS_abs} shows that the proposed scheme can maintain the
SINRs of the M-MSs; however, without the QoS constraint, the introduction
of the femtocell could severely degrade the performance of the M-MSs
especially when the M-MSs are near the F-BS.

\begin{figure}[tbh]
\vspace{-2pt}

\centering \includegraphics[scale=0.36]{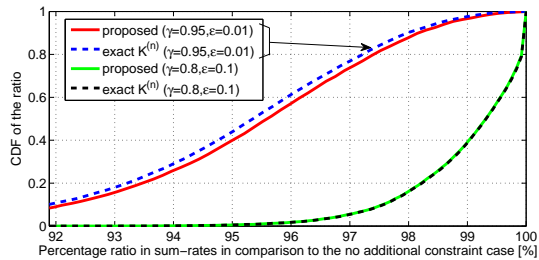}

\caption{Sum-rate degradation of F-MSs.}

\label{fig:FMS}
\end{figure}

Even though the QoS constraint limits the degradation of M-MSs, the
sum-rate of F-MSs becomes worse because the F-BS cannot optimally
allocate its transmit power for the F-MSs with the constraint. The
resulting degradation of the sum-rate of the F-MSs is shown in Fig.
\ref{fig:FMS}. The CDF shows that 10\% of the F-MSs experience 8\%
or more degradation in their sum-rate with the tighter constraint.
With a less tight constraint, only 5\% of F-MSs experience 3\% or
more degradation. The degradation is not significant, especially when
we consider the huge gain that the F-MSs achieve with the introduction
of the femtocell \cite{Kim}. Note that here we do not consider the
degradation caused by periodic unavailability of the frames that are
used for decoding the RAI. More than half of the M-MSs are located
nearer to the M-BS than the F-BS; hence, the approximated value of
$I_{m}$ and consequently that of $K^{(n)}$ for the M-MSs are higher
than the exact ones. Since a higher $K^{(n)}$ leads to a less tight
constraint on the transmit powers of each sub-channel, the sum-rate
of the F-MSs increases whereas that of M-MSs decreases. Therefore,
the proposed scheme is slightly better than the exact one in terms
of the sum-rate of F-MSs.

\section{Conclusions}

We have derived a closed-form solution of the upper limit on the transmission
power of each sub-channel of a F-BS that guarantees the probabilistic
QoS of M-MSs. With the upper limit, the proposed scheme can meet various
QoS constraints of M-MSs even if the M-MSs are very close to the F-BS,
which could be a disaster for those M-MSs without the proposed power
limit. Moreover, compared to the case when a F-BS allocates its power
to maximize only the sum-rate of its own F-MSs, the proposed scheme
avoids significant degradation in the sum-rate while meeting the QoS
constraint.

\appendices{\label{sec:appendix}}

%For the PF scheduling, $F_{X_{PF}}(q)$ is derived as $F_{X_{PF}}(q) = Prob(\max\{x\} \leq q) = (F_X(q))^U = %(q/(1+q))^U$. Then, from the fourth row in eq. (\ref{power_limit}), $K_{PF}^{(n)}$ is represented as
%\begin{eqnarray} \label{PFlimit}
%&&1-F_{X_{PF}} (\kappa^{(n)}/p_F^{(n)}) \leq \epsilon^{(n)}
%~ \Leftrightarrow ~ 1 - \left(\frac{\kappa^{(n)}/p_F^{(n)}}{1+\kappa^{(n)}/p_F^{(n)}}\right)^U \leq \epsilon^{(n)} %\nonumber \\
%&\Leftrightarrow& p_F^{(n)} \leq \kappa^{(n)} ((1-\epsilon^{(n)})^{-1/U}-1) = K^{(n)} \delta^{(n)} %((1-\epsilon^{(n)})^{-1/U}-1)
%\end{eqnarray}

\label{app:KKT} The KKT conditions are as follows:
\begin{eqnarray}
\sum_{n=1}^{N}p_{F}^{(n)}-P_{F}\leq0,~~~\lambda\geq0,~~~\lambda\left(\sum_{n=1}^{N}p_{F}^{(n)}-P_{F}\right)=0\label{KKT_1}\\
-p_{F}^{(n)}\leq0,~~~\mu^{(n)}\geq0,~~~-\mu^{(n)}p_{F}^{(n)}=0\label{KKT_2}\\
p_{F}^{(n)}\leq K^{(n)},~~~\nu^{(n)}\geq0,~~~\nu^{(n)}\left(p_{F}^{(n)}-K^{(n)}\right)=0\label{KKT_3}\\
\frac{-1}{p_{F}^{(n)}+S^{(n)}/\left(H_{F,F_{k}}^{(n)}A_{F}\right)}+\lambda+\nu^{(n)}-\mu^{(n)}=0.\label{KKT_4}
\end{eqnarray}

Two cases follow from (\ref{KKT_3}). For the first case $(p_{F}^{(n)}=K^{(n)},\nu^{(n)}\geq0)$,
the third condition in (\ref{KKT_2}) makes $\mu^{(n)}=0$; from (\ref{KKT_4}),
that equality leads to $\nu^{(n)}=-\lambda+\frac{1}{p_{F}^{(n)}+S^{(n)}/\left(H_{F,F}^{(n)}A_{F}\right)}\geq0$.
Hence, $0<p_{F}^{(n)}=K^{(n)}\leq\frac{1}{\lambda}-\frac{S^{(n)}}{H_{F,F_{k}}^{(n)}A_{F}}.$
For the second case $(p_{F}^{(n)}\leq K^{(n)},\nu^{(n)}=0)$, from
the second condition in (\ref{KKT_2}) and (\ref{KKT_4}), $\mu^{(n)}=\lambda-\frac{1}{p_{F}^{(n)}+S^{(n)}/\left(H_{F,F_{k}}^{(n)}A_{F}\right)}\geq0$.
That is, $\lambda\geq\frac{1}{p_{F}^{(n)}+S^{(n)}/\left(H_{F,F_{k}}^{(n)}A_{F}\right)}.$
If $\lambda<\frac{1}{S^{(n)}/\left(H_{F,F_{k}}^{(n)}A_{F}\right)},$
then $p_{F}^{(n)}>0$; thus, $\mu^{(n)}=0$ according to the third
condition in (\ref{KKT_2}). Consequently, $0<p_{F}^{(n)}=\frac{1}{\lambda}-\frac{S^{(n)}}{\left(H_{F,F_{k}}^{(n)}A_{F}\right)}\leq K^{(n)}$.
Conversely, if $\lambda\geq\frac{1}{S^{(n)}/\left(H_{F,F_{k}}^{(n)}A_{F}\right)},$
then $\lambda\geq\frac{1}{S^{(n)}/\left(H_{F,F}^{(n)}A_{F}\right)}>\frac{1}{S^{(n)}/\left(H_{F,F}^{(n)}A_{F}\right)+p_{F}^{(n)}}\neq0$
assuming that $p_{F}^{(n)}>0$; hence, $\mu^{(n)}>0$, which makes
$p_{F}^{(n)}=0$ according to the third constraint in (\ref{KKT_2}).
Thus, the assumption is contradictory. Hence, $p_{F}^{(n)}=0$ when
$\frac{1}{\lambda}-\frac{S^{(n)}}{\left(H_{F,F_{k}}^{(n)}A_{F}\right)}\leq0$.
Therefore, by incorporating both cases, we get $p_{Fopt}^{(n)}=\min(K^{(n)},\max(0,\frac{1}{\lambda}-\frac{S^{(n)}}{\left(H_{F,F_{k}}^{(n)}A_{F}\right)}))$.

\bibliographystyle{IEEEtran} \phantomsection\addcontentsline{toc}{section}{\refname}

\bibliographystyle{IEEEtran}
\bibliography{IEEEabrv,IEEEexample,NF2_VTC,capacity}

\end{document}